\documentclass[%
 reprint,
superscriptaddress,
twocolumn,
 amsmath,amssymb,
 aps,
prb,
]{revtex4-2}

\usepackage{graphicx}
\usepackage{dcolumn}
\usepackage{bm}
\usepackage{hyperref}

\usepackage[dvipsnames]{xcolor}

\usepackage{soul}

\begin{document}


\title{Collapse of susceptibility and non-trivial spin dynamics in the hyper-honeycomb magnet $\beta$-Li$_2$IrO$_3$ under high pressure}

\author{A. Verrier}
 \affiliation{Institut Quantique, Département de Physique and RQMP, Université de Sherbrooke, 2500 boulevard de l'Université, Sherbrooke, Québec J1K 2R1, Canada}
\author{V. Nagarajan}
\affiliation{Department of Physics, University of California, Berkeley, CA 94720, USA}
\author{L.-T. Gendron}
\affiliation{Institut Quantique, Département de Physique and RQMP, Université de Sherbrooke, 2500 boulevard de l'Université, Sherbrooke, Québec J1K 2R1, Canada}
\author{J. G. Analytis}
\affiliation{Department of Physics, University of California, Berkeley, CA 94720, USA}
\affiliation{Canadian Institute for Advanced Research, Toronto, Canada M5G 1Z8}
\author{J. A. Quilliam}
\email[]{Jeffrey.Quilliam@USherbrooke.ca}
 \affiliation{Institut Quantique, Département de Physique and RQMP, Université de Sherbrooke, 2500 boulevard de l'Université, Sherbrooke, Québec J1K 2R1, Canada}
 \affiliation{Laboratoire de Physique des Solides, Université Paris Sud 11, UMR CNRS 8502, F-91405 Orsay, France}

\date{\today}

\begin{abstract}
We present high-pressure (2 GPa) $^7$Li nuclear magnetic resonance (NMR) measurements on single crystals of the hyper-honeycomb Kitaev magnet $\beta$-Li$_2$IrO$_3$. The spectra show evidence for a structural phase transition around 200 K and a coexistence of phases, consistent with the results of other measurement techniques. The NMR spectra and line shift measurements demonstrate a strong suppression of the local magnetic susceptibility at high pressure. However, the spin-lattice relaxation ($1/T_1$) shows a clear power-law  temperature dependence. This is inconsistent with a gapped singlet ground state of dimers and tetramers, as was previously proposed, and is instead evocative of a more exotic quantum spin liquid-like ground state.
\end{abstract}

\maketitle

\section{Introduction}

Kitaev's discovery of an exactly solvable quantum spin liquid model on the honeycomb lattice~\cite{Kitaev2006} and the subsequent proposal for how his anisotropic bond-dependent model could arise in realistic materials with strong spin-orbit coupling~\cite{Jackeli2009} have led to a revolution in the field of frustrated quantum magnetism and a surge of research into effective spin-1/2 magnets based on $4d$ (eg. Ru) and $5d$ (eg. Ir) orbitals. This has led to discoveries in honeycomb systems like $\alpha$-RuCl$_3$ where continua in inelastic neutron scattering~\cite{Banerjee2017} and possible quantized thermal Hall conductivity~\cite{Kasahara2018} provide tantalizing indications of quantum spin liquidity. While these results are not without controversy~\cite{Winter2017,Czajka2022,Lefrancois2022}, there is little doubt that strong spin-orbit coupling and an important Kitaev contribution to the exchange Hamiltonian are giving rise to a great deal of interesting physics. 

Furthermore, it has been realized that the anisotropic Kitaev model is highly frustrated on a wide variety of lattices~\cite{Trebst2022,OBrien2016}, several of which are realized in real materials, notably the triangular lattice of Ba$_3$TiIr$_2$O$_9$~\cite{Lee2017}, the stripy-honeycomb lattice of $\gamma$-Li$_2$IrO$_3$~\cite{Modic2014} and, the subject of this work, the hyper-honeycomb lattice, which is approximated by $\beta$-Li$_2$IrO$_3$~\cite{Takayama2015} and $\beta$-ZnIrO$_3$~\cite{Haraguchi2022}. Mandal and Surendran showed that Kitaev's hamiltonian could be defined on a hyperhoneycomb lattice and exactly solved for homogeneous $K_\gamma$~\cite{Mandal2009}. Similarly to the honeycomb case, this solution features a quantum spin liquid ground state with Majorana Fermions that are either gapped or gapless, depending on the values of the three possible $K_\gamma$ interaction strengths. 

The hyperhoneycomb material $\beta$-Li$_2$IrO$_3$ is unfortunately not a spin liquid at ambient pressure. It was found to have net ferromagnetic interactions ($\theta_W~=~+40$~K) but an antiferromagnetic ground state below $T_N~=~38$~K~\cite{Takayama2015, Biffin2014,Ruiz2017}, suggesting competition between a ferromagnetic Kitaev interaction and an antiferromagnetic Heisenberg interaction~\cite{Takayama2015}. The magnetic structure is found to be a particularly complicated incommensurate order with non-coplanar and counter-rotating moments~\cite{Biffin2014,Ruiz2017}. A subtle bulk magnetic phase transition is also observed at 100~K but is easily suppressed with magnetic fields 0.5~T or larger~\cite{Ruiz2020}.

 \emph{Ab initio} calculations~\cite{Kim2015} were used to determine a more general spin model 
\begin{equation}
\mathcal{H}_{ij}~=~J\vec{S}_i\cdot\vec{S}_j + KS_i^\gamma S_j^\gamma + \Gamma \left( S_i^\alpha S_j^\beta + S_i^\beta S_j^\alpha \right) 
\end{equation}
with dominant Kitaev interaction ($K$), but non-negligible Heisenberg exchange ($J$) and symmetric bond-dependent exchange ($\Gamma$) that is found to be consistent with the experimentally observed magnetic structure~\cite{Lee2015}. Despite the presence of magnetic order, indications of fractionalized excitations in Raman spectroscopy have been reported~\cite{Glamazda2016}. 

Moreover, a collapse of the magnetic moments under modest pressures $\gtrsim 1.4$~GPa was quickly observed with X-ray magnetic circular dichroism (XMCD)~\cite{Takayama2015} and later confirmed with magnetic susceptibility and muon spin rotation ($\mu$SR)~\cite{Majumder2018}. It is tempting to attribute such a suppression of magnetism to a pressure-induced quantum spin liquid phase, especially given that further DFT calculations show significant changes to the spin model under pressure~\cite{Kim2016}. It is seen that at high enough pressures it is the $\Gamma$-term that becomes dominant. It was pointed out~\cite{Kim2016} that this term also gives rise to frustration on the hyper-honeycomb lattice and might provide a mechanism to generate a novel spin liquid state, whose precise nature has not yet been studied.

That said, the effects of pressure on the material's structure are far from trivial. At room temperature, $P > 4.05$~GPa is required to induce a structural phase transition from an $Fddd$ structure to a high pressure $C2/c$ phase~\cite{Veiga2017} indicating that the collapse of magnetism at $P \simeq 1.4$~GPa was not associated with a change in structure. Instead, X-ray absorption near-edge structure (XANES) imply that there is a change in spin-model associated with a reduction of the strength of spin-orbit coupling under pressure~\cite{Veiga2017}. However, subsequent measurements showed that additional structural phases are induced under much lower pressures at low temperatures~\cite{Veiga2019}. At $T \leq 25$~K, a coexistence of $Fddd$ and $C2/c$ phases was evidenced between 1.5 and 2.8~GPa, although an unambiguous refinement could not be achieved~\cite{Veiga2019}. At 50~K another intermediate pressure phase with $P2_1/n$ symmetry was also observed. While the highest pressure phase showed clear evidence for dimerization (shortening of the $\gamma = Z$ Ir-Ir bonds), it is much less clear what is occurring in the intermediate pressure coexistence phase, where $Fddd$ $Z$-bonds are shortened and $C2/c$ $Y$-bonds are shortened. 

The most recent study of this material has outlined its structural and magnetic phase diagram using magnetic susceptibility measurements, confirming that a first structural phase transition occurs at roughly 1.4~GPa when temperatures are below 150~K~\cite{Shen2021}. In that same work, \emph{ab initio} structure calculations are employed and reveal a transition to a partially dimerized $P2_1/n$ phase at 1.7~GPa, followed by a transition to a fully dimerized $C2/c$ phase at 2.7~GPa. This is somewhat consistent with the X-ray diffraction results of Veiga \emph{et al.}~\cite{Veiga2019} that were obtained at 50~K, but not the coexistence of phases $Fddd$ and $C2/c$ at lower temperatures. The \emph{ab initio} calculations also allow for an estimation of a magnetic model in the partially dimerized phase and it was determined that the system is broken up into strong Ir2-Ir4 dimers and Ir1-Ir3-Ir3-Ir1 tetramers.

From a theoretical perspective, the system has a trivial singlet ground state with a 9~meV gap on the tetramers~\cite{Shen2021}. However, this is much less clear from an experimental perspective. A small Curie-Weiss term at low temperatures is measured in the partially dimerized phase which has been attributed to impurities, but the idea of a cluster magnet of tetramers has also been invoked~\cite{Shen2021}. Additionally, $\mu$SR measurements at around 2~GPa only provide evidence of a partial suppression of magnetic freezing, implying that there is a significant volume fraction of the sample that is not in a singlet ground state~\cite{Majumder2018}. 

To resolve this puzzle, we have carried out nuclear magnetic resonance (NMR) measurements under pressure. As a local, site-selective probe, NMR can in principle provide information on what volume fractions, and even what Ir sites in the sample, participate in magnetic freezing, spin liquid-like correlations or singlet formation. Our measurements confirm that at a pressure of 2~GPa, significant structural and/or magnetic changes occur. Notably we find that the pressure-induced phase, while indeed having a drastically reduced magnetic susceptibility, is not consistent with a simple gapped singlet state.

\begin{figure*}[hbtp]
	\centering
	\includegraphics[height=5in]{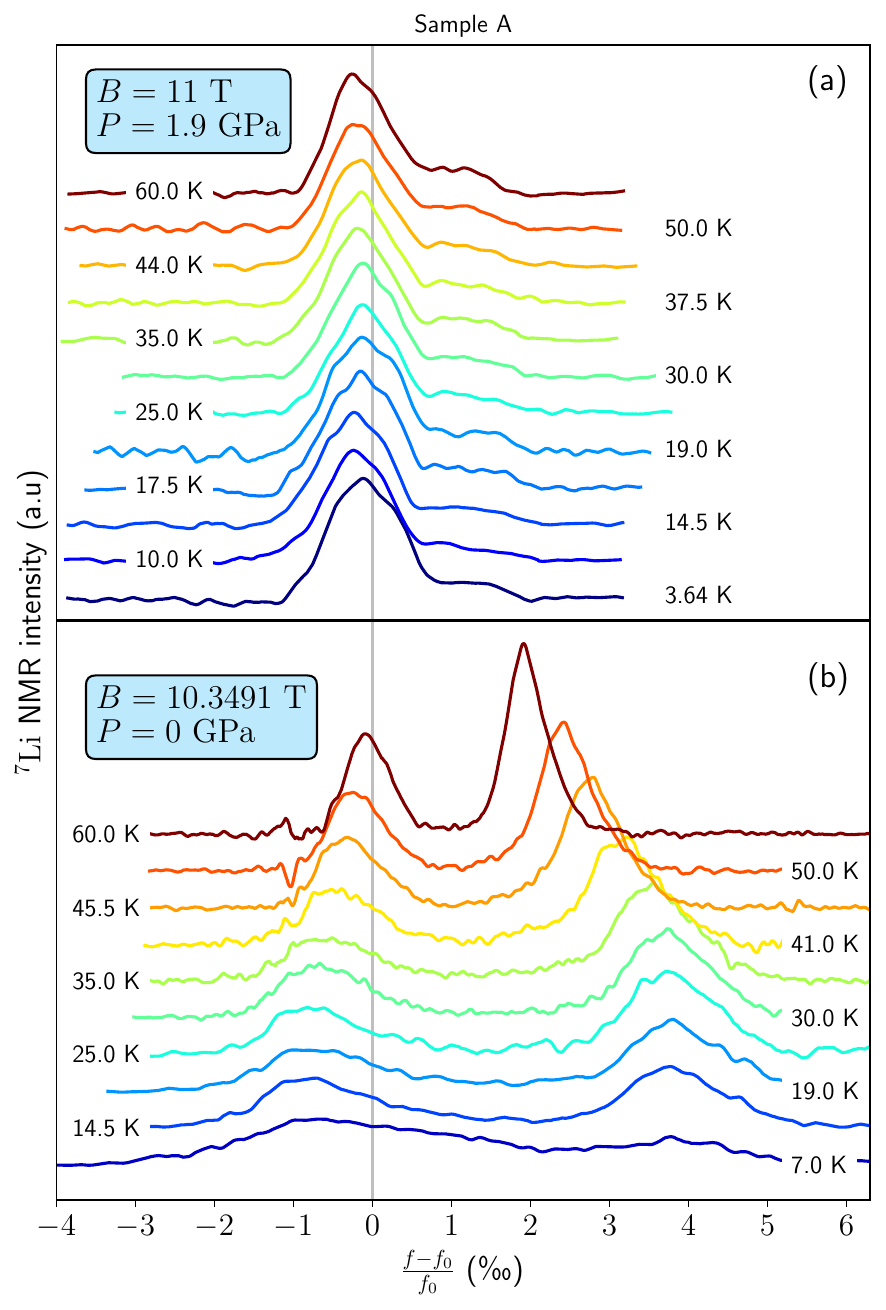}
	\includegraphics[height=5in]{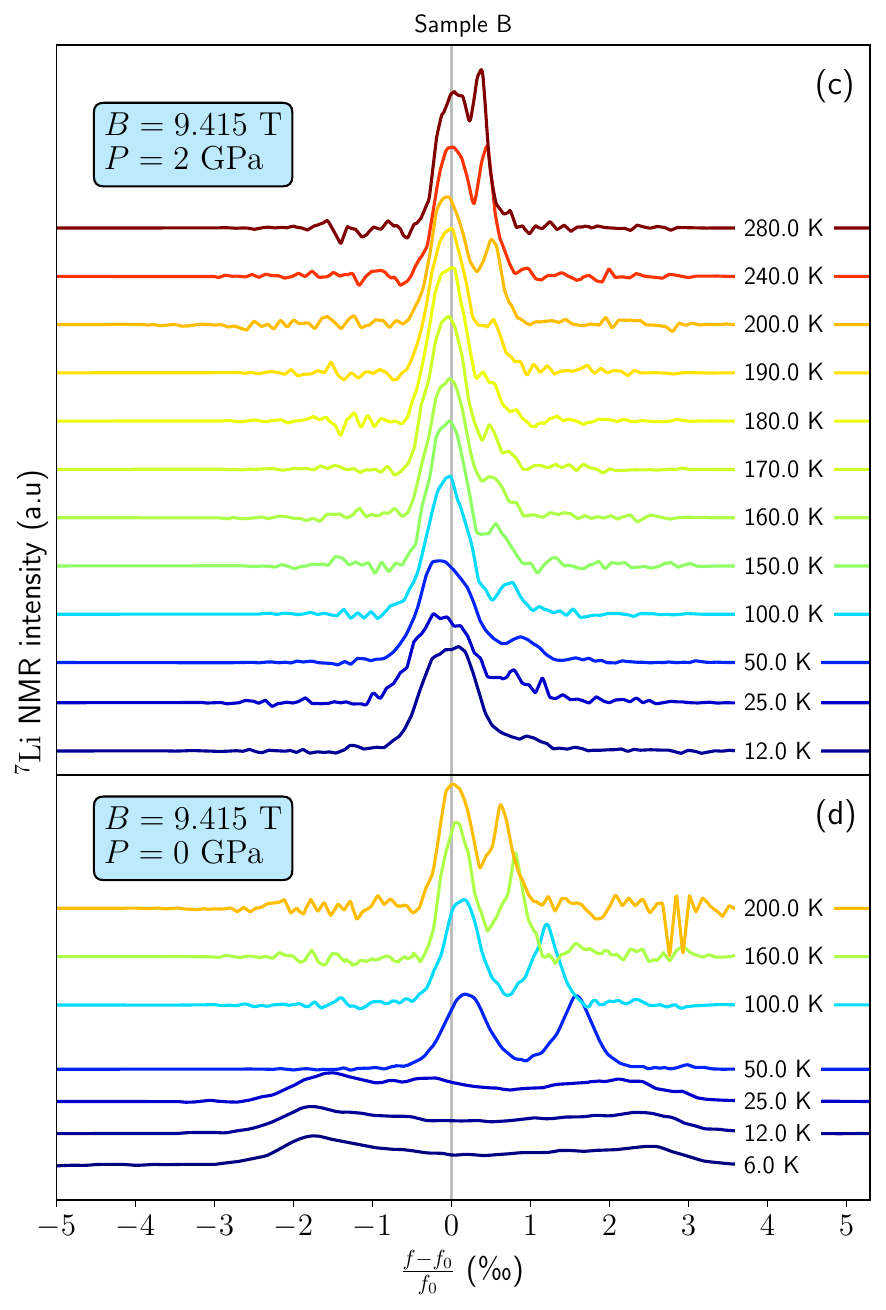}
	\caption{$^7$Li NMR spectra as a function of temperature at low and high pressure for both Samples. Aqueous LiCl at 280~K is the reference for $K=0$. Somewhat different magnetic fields were used for the high and low pressure measurements, as indicated on the graph. }
	\label{fig:spectra}
\end{figure*}

\section{Experiment}

Single crystals of $\beta$-$\mathrm{Li}_2\mathrm{IrO}3$ were synthesized using an isothermal vapor transport technique described in \cite{Ruiz2017}. Ir (99.9\% purity, BASF) and Li2CO3 (99.999\% purity, Alfa-Aesar) powders were ground in the molar ratio of 1:1.05, then pressed into a pellet which was reacted at 1050 $^\circ$C for 12 h, and then cooled down to 850 $^\circ$C at 2 $^\circ$C/h. Two of such crystals (hereafter referred to as sample A and sample B) of size on the order of $0.2~\mathrm{mm}\times0.1~\mathrm{mm}\times0.1~\mathrm{mm}$, were aligned and glued with epoxy in separate copper coils. Each coil and its contents was, in turn, aligned and inserted into a self-clamped hydrostatic BeCu pressure cell with a 5~mm bore along with a lead pressure gauge, using Daphne 7373 as pressure medium. The pressure was increased and measured at room temperature through the variation of resistance of the lead gauge compared with a calibration curve~\cite{Eiling1981}. To make sure that the pressure did not change significantly between high and low temperature, the superconducting critical temperature of the lead gauge accompanying sample B was measured before and after pressurization and combined with $\partial T_\mathrm{c}/\partial P~=~-0.361~\pm~0.005~\mathrm{K}/\mathrm{GPa}$ following Ref.~\cite{Clark1978} to infer the pressure at low temperature. This pressure cell configuration was previously tested using the nuclear quadrupole resonance (NQR) frequency of Cu$_2$O, suspended in the same epoxy, as has been demonstrated in several prior works~\cite{Fukazawa2007,Hirayama2008,Kitagawa2010}. The line width of the NQR signal indicated that the homogeneity of the pressure was better than 2~\%.

NMR $^7\mathrm{Li}$ spectra and spin-lattice relaxation rate $1/T_1$ were all measured in zero-field-cooled conditions in order to avoid possible hysteresis effects since there is long range order at zero pressure~\cite{Biffin2014}. Every individual spectrum was acquired at fixed magnetic field $\vec{B}\parallel\vec{c}$ and fixed \textit{RLC} resonance by scanning the carrier frequency of a Tecmag Redstone spectrometer and reconstructing the spectrum as described in \cite{Verrier2020}. 
The metallic $^{65}\mathrm{Cu}$ signal from the coil was used as an \emph{in situ} reference and was calibrated to an acqueous LiCl solution at room temperature. Large fields of $\sim$10~T were used so that the signal would be of reasonable quality despite the small size of the sample. To measure $T_1$, a $[\pi/2-\tau_2]_{10\times}-\mathcal{T}-\pi/2-\tau-\pi$ (saturation recovery) sequence was used and the signal was fitted with a stretched exponential relaxation
\begin{equation}
M(\mathcal{T})=M_{\mathrm{eq}}\left(1-f_{\mathrm{sat}}e^{-\left(\mathcal{T}/T_1\right)^\beta}\right)
\end{equation}
where $\beta$ is the stretching exponent, which accounts for a distribution of relaxation rates.


Coarse preliminary measures of $T_2$ and $T_1$ allowed us to determine values of $\tau$, $\tau_2$ and repetition rate that would optimize SNR per unit of time while avoiding distortions of the spectra. As an example, $T_2\approx190(460)~\mu s$ for Li1(Li2) at 120~K and 0~GPa or 50(170) at 37.5~K and 2~GPa.

The magnetic field was aligned as closely as possible along the $c$-axis of the crystal. This orientation, which is perpendicular to the material's easy ($b$) axis~\cite{Majumder2019}, was chosen so as to remain in the magnetically ordered phase (at ambient pressure) even in relatively high magnetic fields. Magnetic susceptibility and specific heat measurements~\cite{Ruiz2017,Majumder2019, Majumder2020}, have shown that a magnetic field of only 3~T along $\hat{b}$ causes the system to leave the antiferromagnetic phase and end up in a largely field-polarized paramagnetic phase. Our aim is to use sufficiently high magnetic fields in order to work with a single-crystal sample, but avoid the trivial field-polarized phase. Previous NMR measurements were carried out on either a mosaic of single crystals co-aligned just along the $b$-axis~\cite{Majumder2019} or else a polycrystalline sample~\cite{Majumder2020}. In order to avoid gluing samples to a sample support and possibly compromising the homogeneity of the applied pressure, we have chosen to work with a single crystal at a time. While greatly reducing the NMR signal, it has the advantage of significantly increasing our resolution as compared to previous work~\cite{Majumder2019,Majumder2020}. In order to achieve the most direct comparison possible, the ambient pressure experiments of both samples were carried out with the sample inside of the pressure cell prior to pressurization.

\section{Results}

Two separate single-crystal samples of $\beta$-Li$_2$IrO$_3$ (sample A and sample B) were measured in order to verify that our results were reproducible and, in the case of Sample B, to cover a larger range of temperatures. Spectra of both samples at ambient and high pressure are shown in Fig.~\ref{fig:spectra}. In the bottom panels~\ref{fig:spectra}(c) and~\ref{fig:spectra}(d) spectra taken at ambient pressure are shown, whereas high-pressure measurements are shown in the top panels~\ref{fig:spectra}(a) and~\ref{fig:spectra}(b).

\begin{figure*}[htbp]
	\centering
	\includegraphics[width=\linewidth]{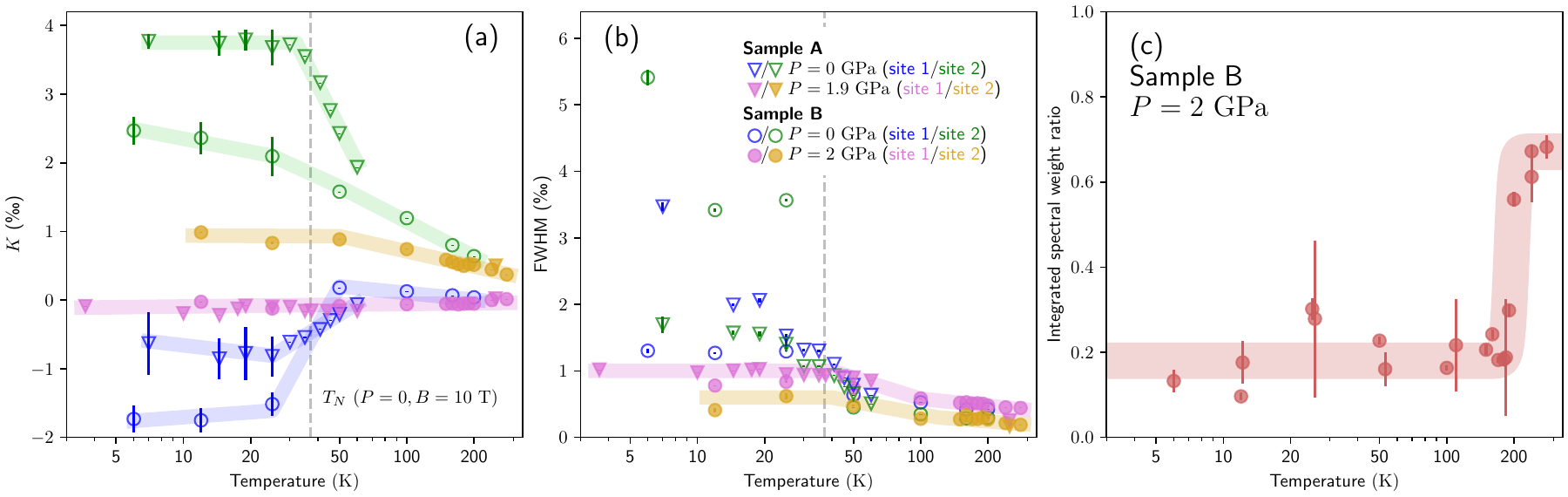}
	\caption{Features of the $^7$Li NMR spectra as a function of temperature. Shift (a) and spectral width (b) were obtained as explained in the text. Triangles represent properties of Sample A and circles represent properties of Sample B.  Ambient pressure measurements are given by open blue (Li1) and green (Li2) symbols. High pressure measurements are represented with closed magenta (Li1) and orange (Li2) symbols.The integrated spectral weight ratio (c) is the area of the fitted Li2 curve relative to Li1.}
	\label{fig:shift}
\end{figure*}

\subsection{Ambient pressure spectra}

As can be seen in Figs.~\ref{fig:spectra}(b) and (d), the ambient pressure spectra of both samples show two distinct peaks which have been identified as arising from the two distinct Li sites in the crystal structure. Following Majumder \emph{et al.}~\cite{Majumder2019} we label the site that has a negative coupling (negative shift) as Li1 and the site with a positive, and larger, shift as Li2. With decreasing temperature, the magnetic susceptibility increases leading to a larger separation between the two peaks. As observed previously~\cite{Majumder2019}, the nuclear quadrupolar splitting appears to be small enough that we only observe a single line per crystallographic site, presumably due to the almost regular oxygen octahedra that surround both Li sites.

Values of shift ($K$) and line width extracted from the spectra are shown in Figs.~\ref{fig:shift}(a) and (b), respectively. For relatively high temperatures the shift and line widths have been obtained by fitting the spectra with two peaks corresponding to the two inequivalent Li sites. A Lorentzian line shape provided a better fit to the Li2 peak, whereas a Gaussian line was used for the Li1 peak. We speculate that the quadrupolar satellites are so close to the central peaks that they are impossible to distinguish, yet nonetheless have an effect on the line shape. Below 30~K, it was necessary to use an asymmetric line shape, in particular a skew normal distribution~\cite{OHagan1976}, to fit the data. The line width was taken from the full-width at half maximum (FWHM) of the fitted function and the peak position was simply extracted from the local maximum of the data, to avoid a bias in the shift due to overlap of asymmetric peaks. Relative uncertainty in the shift at low $T$ was estimated from the width at 90\% of the peak height, a threshold which was chosen based on the noise in the data. At higher temperatures the smaller error bars are derived from the uncertainty in the fitted parameters. The extracted line width is not entirely reliable since the spectral weight in the middle of the spectrum is ambiguously attributed to either of the peaks. Nonetheless, it provides a qualitative indication of the broadening of the spectra.

Down to the Néel temperature, the shift $K(T)$ ought to provide a measurement of the local magnetic susceptibility. Above 100~K, the $b$-axis and $c$-axis susceptibilities are not very different~\cite{Majumder2019} and from that temperature range we can conclude that the Li2 site has essentially the same hyperfine coupling as the one obtained from the data in Majumder \emph{et al.}~\cite{Majumder2019}, that is $A_2~=~1.8$~kOe/$\mu_B$~\footnote{Note that there appears to be a small unit-conversion error in Ref.~\cite{Majumder2019} that has led the authors to report values of hyperfine coupling a factor of 10 too small.}. In other words $A_2$ appears to be quite isotropic. In this configuration, we lacked the precision to extract the smaller hyperfine coupling of the the Li1 site, so we assume the value of $A_1~=~0.47$~kOe/$\mu_B$ that can be extracted from the data in Ref.~\cite{Majumder2019}, to be correct and isotropic. 

Comparing the two samples' ambient-pressure spectra in Fig.~\ref{fig:spectra}(b) and (d) reveals some differences. Notably Sample B has overall narrower spectra, with a smaller shift of the Li2 site and a slightly larger Li1 shift at low temperatures. This difference is also evident in the values of shift extracted from the spectra which are plotted as a function of temperature in Fig.~\ref{fig:shift}(a) (open symbols corresponding to the ambient pressure measurements). We attribute this quantitative difference in shift to the strongly anisotropic $g$-tensor~\cite{Majumder2019} and slightly different crystal alignment with respect to the magnetic field, which is difficult to avoid within the pressure cell. Taking the highly anisotropic susceptibilities measured at low temperatures ($\chi_{yy}~\gg~\chi_{zz}$)~\cite{Majumder2019} and an isotropic hyperfine constant, we estimate that Sample A is misaligned by at most 16$^\circ$ and Sample B at most 6$^\circ$.

Biffin \emph{et al.}~\cite{Biffin2014} have determined the magnetic structure to be a complicated incommensurate, counter-rotating and non-coplanar spin spiral. In NMR, even on a single crystal, incommensurate order is expected to broaden the lines with a characteristic ``two-horn'' pattern~\cite{Sakhratov2016, Ding2017}. While we observe line broadening below $T_N$, as indicated in Fig.~\ref{fig:shift}(b), we see no evidence of the two-horn incommensurate line shape. This is also the case in previous single-crystal and powder NMR results~\cite{Majumder2019,Majumder2020}. Measurements of a mosaic of crystals with the field applied in the $ac$-plane~\cite{Majumder2019} and our measurements show a quantitatively similar level of field-broadening ($\Delta H$) below the Néel temperature.  Furthermore, the asymmetrical shape of the lines at $T\ll T_N$ implies that the applied magnetic field has a significant influence on the spin structure and that the $c$-axis magnetization cannot be modeled with a simple expression that sums homogeneous field-induced magnetization and incommensurate antiferromagnetism, i.e. $m_z(\vec{r}) = m_{z0} + S_{z0} \cos(qx)$~\cite{IncommensurateFootnote}

Nonetheless, we can compare the observed line width with what should be expected from the antiferromagnetic structure and magnetic moment of 0.47~$\mu_B$/Ir that were determined by Biffin \emph{et al.}~\cite{Biffin2014} in zero magnetic field. To estimate the expected line width of the Li2 peak, we assume that the hyperfine constant $A_2$ is equivalent for the 5 closest Ir sites. Averaging the contributions from those 5 Ir sites for a Li2 site at position $x$ along the $a$-axis gives $S_{z,\mathrm{av}} = S^{z0} \cos(qx) [4\cos(q/4) + 1]/5$ where $q~=~0.57(2\pi)$ is the propagation vector of the magnetic structure~\cite{Biffin2014} and $x$ is in normalized coordinates. This leads to an oscillating $z$-component of the internal magnetic field at the Li2 sites with an amplitude of $A_2 S_{z,\mathrm{av}}~=~0.059$~T. Thus we should expect a two-horn pattern (for the Li2 site) with a full width of 0.118~T or roughly 1.2~\% of the applied field. Here, the standard incommensurate spectrum seems to be smeared out, giving a somewhat smaller line width of $\sim 0.55$\% as shown in Fig.~\ref{fig:shift}(b). Ruiz \emph{et al.}~\cite{Ruiz2017} have already observed that an applied magnetic field with a component along the $b$-axis quickly reduces the size of the order parameter, which may explain the smaller line width observed here. In any case, our ambient-pressure measurements are broadly consistent with previous NMR measurements~\cite{Majumder2019,Majumder2020} and show evidence of magnetic order with a moment size that is a significant fraction of that found in Ref.~\cite{Biffin2014}.

\subsection{High pressure spectra}

We now consider the high pressure (2~GPa) measurements, which should have sample orientations identical to the ambient pressure measurements. As can be seen from Figs.~\ref{fig:spectra}(a) and (b), a dramatic change in the spectra occurs under applied pressure. In the case of Sample A, almost the entirety of the spectral weight is within $\pm~0.5$~\textperthousand~of the reference. A small shoulder is barely perceptible around $\simeq~+1.2$~\textperthousand. As is expected from the bulk magnetization measurements~\cite{Majumder2018}, there is evidently a significant collapse of the magnetic susceptibility, and therefore Knight shift, $K$. 

Very similar results are obtained for Sample B, Fig.~\ref{fig:spectra}(c), although the spectra are overall slightly narrower. The shoulder at positive shift becomes more apparent in Sample B than in Sample A as we look at higher temperatures. With Sample B we have also taken spectra at higher temperatures and here we see evidence of the pressure-induced phase transition that was identified in previous works~\cite{Veiga2019,Shen2021}. At 280~K, above the structural phase transition, the spectrum appears to consist of two peaks with similar weight (Li2 has an intensity roughly 70~\% of that of Li1). At lower temperatures, the integrated spectral weight ratio $I_2/I_1$, as shown in Fig.~\ref{fig:shift}(c), drops to about 20~\%. The transition between these two regimes occurs between 190 and 200~K and is almost certainly a consequence of the structural phase transition observed by Shen \emph{et al.}~\cite{Shen2021}. This intensity ratio has not been corrected for differences in $T_2$ between the two sites, thus it should not be interpreted as an accurate measure of the volume ratio of the two phases. The reduced spectral weight of the Li2 site is likely due to the collapse of magnetic susceptibility in a large fraction of the sample, which eliminates the otherwise significant shift of this site and causes it to be superimposed on the Li1 site. The overall intensity stays more or less the same with and without applied pressure.

Judging by the width of the main peak under pressure and the difference in shift between the two peaks, the magnetic susceptibility must be abruptly reduced by at least a factor of 2 or 3 at the transition. At lower temperatures, this difference becomes much more significant, a factor of 6 or more. Shen \emph{et al.}~\cite{Shen2021} observe what appears to be a more subtle drop in susceptibility at the structural transition, but they also note that they have an imperfect background subtraction, which is not an issue in NMR measurements.

The fact that the distinct Li2 site maintains a small but finite spectral weight down to low temperatures implies that there is a coexistence of phases in the material as was proposed by Veiga \emph{et al.}~\cite{Veiga2019}. The minority phase evidently retains a fair degree of magnetism with a shift of around 1~\textperthousand~at low temperatures, as can be seen in Fig.~\ref{fig:shift} by following the shift of the filled symbols, which is roughly 40~\% of the ambient-pressure shift for the same sample.

\begin{figure*}[htbp]
	\centering
	\includegraphics[width=\linewidth]{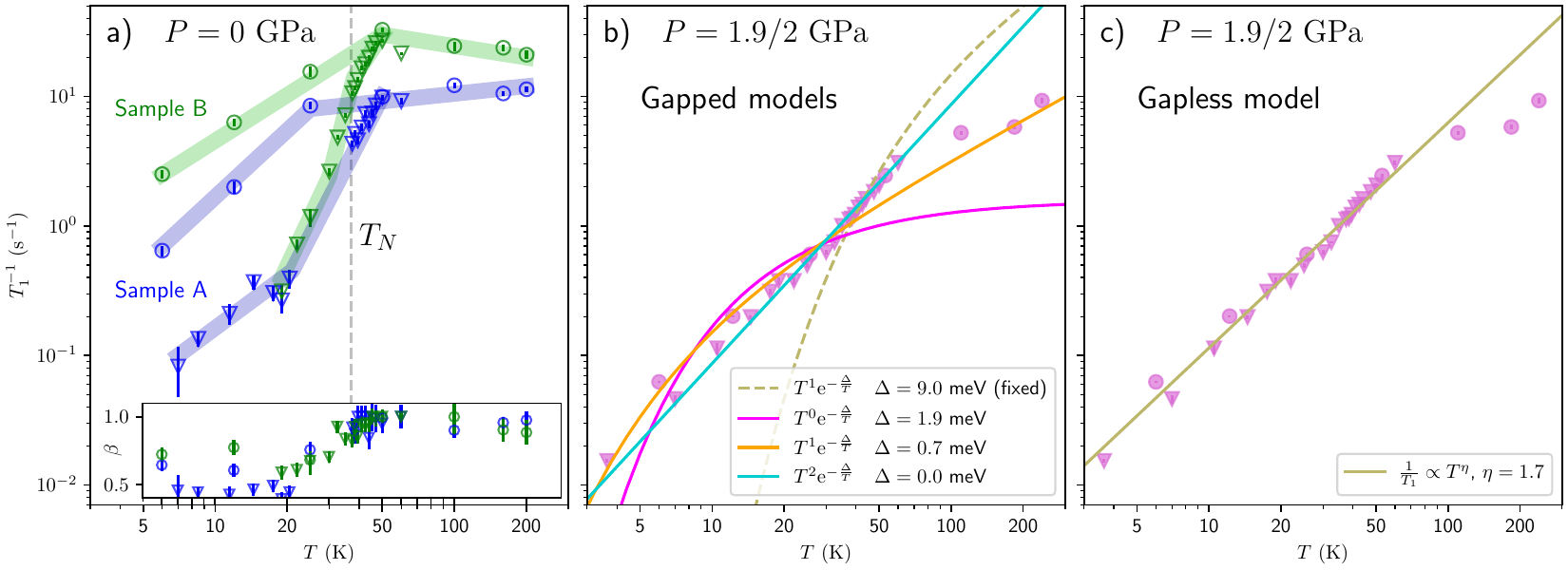}
	\caption{(a) Variation of relaxation rate with temperature at ambient pressure. The temperature-dependent stretching exponent obtained from fits to the recovery curves is shown in the inset. Shaded areas are guides to the eye. (b) High pressure relaxation rate on the main line for both samples. The solid lines are fitted to the data under 100~K allowing the gap to vary whereas the dashed line uses a fixed gap of 9~meV as proposed by Shen \emph{et al.}~\cite{Shen2021} (c) The same data obtained under pressure are fitted with a gapless power law. This shows that the relaxation is incompatible with a large spin gap. Symbol formatting in the entire figure follows the same convention as Fig.~\ref{fig:shift}.}
 	\label{fig:relaxation_rate}
\end{figure*}

The scenario of Shen \emph{et al.}~\cite{Shen2021}, wherein there is a single $P2_1/n$ structural phase with a mix of dimers and tetramers  (as opposed to a coexistence of phases), is more difficult to reconcile with our data.  One possible way to do so is to consider changes in the hyperfine coupling constants for inequivalent Li sites. The $P2_1/n$ phase has a lower symmetry than the $Fddd$ phase and thus has 8 inequivalent Li sites (as opposed to 2 at low pressure). It is not impossible that 6 of those 8 sites have a rather small hyperfine coupling and are all superimposed near $K=0$ while the two remaining sites have a stronger hyperfine coupling and give rise to the small peak at positive shift. This picture would give rise to a similar spectral weight ratio to that which is measured in Fig.~\ref{fig:shift}(c). However, since the lattice parameters and interatomic distances evolve rather gradually between the various phases~\cite{Veiga2019}, we find it highly unlikely that 2 of the 8 hyperfine constants would experience such dramatic changes at the transition.

It is similarly unlikely that different Li sites are more or less strongly coupled to the non-magnetic dimers and magnetic tetramers in Shen \emph{et al.}'s scenario. All of the Li sites are about equidistant from the Ir2/Ir4 sites (which are proposed to participate in the dimers) and the Ir1/Ir3 sites (which ostensibly make up the tetramers). Therefore each $^7$Li nuclear spin should simultaneously be sensitive to the dimers and tetramers. The dimer-tetramer scenario therefore likely cannot account for the small peak at positive shift and we conclude that the most reasonable explanation is a coexistence of phases.

It is also important to note that we do not see any significant line broadening or drop in NMR intensity (i.e. a wipe-out, such as observed in Ref.~\cite{Imai2021} for instance) around $\sim 15$ K, in either of the fractions of our sample, that might indicate spin freezing as observed in the $\mu$SR measurements of Ref.~\cite{Majumder2018}. That said, we do not expect that our measurements performed at around 10 T can have the same level of sensitivity to weakly frozen spins as zero-field $\mu$SR measurements.

\subsection{Ambient pressure relaxation}

In order to better understand the origin of the suppression of magnetic susceptibility in these intermediate pressure phases, we have turned to relaxation rate measurements, $1/T_1$. At ambient pressure, Fig.~\ref{fig:relaxation_rate}(a), we find results that are qualitatively similar to previous measurements by Majumder \emph{et al.}~\cite{Majumder2020} which were nonetheless carried out on powder samples at much lower magnetic fields. For both samples, we observe a very broad maximum in $1/T_1$ in the vicinity of $T_N$. At this point there is also a gradual change in the stretching exponent $\beta$ used to fit the recovery curves which is shown in the inset of Fig.~\ref{fig:relaxation_rate}(a). Unlike  $1/T_1$ measurements reported in Ref.~\cite{Majumder2019}, we do not observe a sharp peak at $T_N$.

Below $T_N$, as seen in Fig.~\ref{fig:relaxation_rate}(a), $1/T_1$ of Sample A drops out much more quickly than that of Sample B. This difference is likely a consequence of differences in sample orientation. As indicated by the shift measurements, sample B appears to be more closely oriented along the $c$-axis than Sample A, which likely has a non-negligible $b$-axis component to the magnetic field. The effect of the magnetic field on the sample is much stronger along the easy $b$-axis. Majumder \emph{et al.}~\cite{Majumder2020} have seen in powder samples that for magnetic field $<~3$~T, the relaxation rate largely plateaus below the Néel temperature and does not drop out appreciably until the temperature is lowered below 2~K. As the magnetic field is increased, the relaxation rate drops at higher temperatures and with a steeper temperature dependence. 

Our results on Sample B are quite similar to the powder results of Ref.~\cite{Majumder2020} for fields close to 5~T. Given that we have applied a field of 9.4~T, it seems likely that the field is fairly closely aligned with the hard axis ($c$-axis), so that its effects are relatively small. On the other hand, for Sample A, the relaxation rate drops out more quickly than the data of Ref.~\cite{Majumder2020} which are only measured up to 5.14~T. Again, this indicates that sample A sees a larger component of the field along $\hat{b}$, leading to a larger effect of the magnetic field. 

\subsection{High pressure relaxation}

Finally we consider the relaxation rate of the main peak under applied pressure (which primarily represents the phase in which the magnetic susceptibility is highly suppressed) which is shown in Figs.~\ref{fig:relaxation_rate}(b) and (c). Under pressure, the relaxation rate at high temperature is similar to the ambient pressure value. In contrast to the sharp suppression of $\chi$ observed at the structural transition at $T_s~\simeq~200$~K ({ and the associated loss of spectral weight of the Li2 peak), $1/T_1$ drops gradually as a function of temperature. It does not go through a maximum around $T_N$ and the recovery curves do not show any significant changes in exponent $\beta$ (which was therefore fixed to 0.8 in the fitting routine). 

From about 100~K down to the lowest temperatures studied, a gapless power-law fit $1/T_1~\propto~T^\eta$ with $\eta~=~1.7$ can be very successfully applied to the data. Nearly identical values of $1/T_1$ are obtained for both samples studied, and the power-law fit shown in Fig.~\ref{fig:relaxation_rate}(c) has been applied to the union of both data sets.

This surprising result suggests that appreciable magnetic fluctuations remain active under pressure despite the apparent suppression of magnetic susceptibility. Since $1/T_1~\propto~\chi''(\omega_0)$, one would expect the observed  drop in susceptibility to lead to a proportional drop in $1/T_1$. However, a drop in moment size may also lead to a reduction in the interaction strength between moments, thereby slowing down fluctuations and increasing their spectral weight at the Larmor frequency $\omega_0$, which would result in a much more subtle change in $1/T_1$. 

Evidently this power law is incompatible with a gapped (singlet) ground state. In Fig \ref{fig:relaxation_rate}(b) we show our best attempts to fit the data to various gapped models. In order to place a conservative limit on the size of a possible gap, we have carried out fits of the form $1/T_1 \propto T^\eta \exp(-\Delta/T)$ with several possible values of the pre-factor exponent $\eta$. The appropriate value of $\eta$ depends on the dispersion relation and dimensionality of the magnetic excitations above the gap energy, as discussed at length in Ref.~\cite{Jansa2018}. For instance, Fu \emph{et al.}~\cite{Fu2015} have used $\eta = 2$ in their analysis of the $^{17}$O relaxation rate in the kagome QSL candidate herbertsmisthite and Jansa \emph{et al.}~\cite{Jansa2018} have used $\eta = -1$ in the candidate Kitaev QSL $\alpha$-RuCl$_3$. Here we can see that $\eta = 1$ (the orange curve) or $\eta = 2$ (the light blue curve) provide relatively good fits to the data, but for $\eta = 1$, we obtain a rather small value of the gap of $\Delta = 0.7$ m$e$V and for $\eta = 2$ the gap energy converges to zero. Since 0.7 m$e$V$/k_B = 8$ K, which is not very far from our lowest temperature data point, we can see that if there is a gap, it is small enough that it has very little impact on our measurements. Using a higher power law fit, $\eta = 2$ (the magenta curve), gives a larger value of $\Delta = 2.0$ m$e$V, but a very poor fit to the data. Leaving the exponent $\eta$ free to vary, the fit will simply converge to $\Delta = 0$ giving the aforementioned power-law with $\eta = 1.7$.

Hence we can place a conservative upper bound on the size of the spin gap in this system at 0.7 m$e$V (obtained with an exponent $\eta = 1$). Lower temperature measurements would be required to place a stricter constraint on the value of a possible gap. The much higher gap of 9~meV which was proposed by Shen \emph{et al.}~\cite{Shen2021}, given by the khaki dashed line in Fig.~\ref{fig:relaxation_rate}(b), is completely incompatible with the data.

\section{Discussion}

Our high-pressure (2~GPa) NMR results show clear evidence of the same structural phase transition and coexistence of phases that was obtained in Ref.~\cite{Veiga2019}. This phase transition is characterized by a significant suppression of the magnetic susceptibility in the majority fraction of the sample and a partial reduction of the magnetic susceptibility in the minority fraction. The observed coexistence of phases is consistent with the conclusions of the high-pressure X-ray diffraction study of Veiga \emph{et al.}~\cite{Veiga2019} who find at low temperatures and intermediate pressures (in the range of around 1.5-2.5~GPa) a coexistence of the low-pressure $Fddd$ phase and the high-pressure $C2/c$ phase. We assume that  the majority phase we observe in our data, in which $K$ is heavily suppressed, corresponds to the $C2/c$ phase whereas the $Fddd$ phase retains about 40~\% of the magnetic moment. This can be seen from the ratio of the Li2 shift with and without pressure as shown in Fig.~\ref{fig:shift}. 

With $\mu$SR measurements, Majumder \emph{et al.}~\cite{Majumder2018} have also shown evidence of a coexistence of phases between 1.36 and 2.27~GPa. About 60~\% of their sample shows dynamic spin fluctuations and a complete lack of spin freezing or magnetic order. The remaining 40~\% shows weak spin freezing below around 20~K. The majority fraction observed in our NMR data clearly shows no evidence of spin freezing and is consistent with the majority fraction seen in $\mu$SR~\cite{Majumder2018} which remains dynamic down to very low temperatures. However, the observation of spin freezing in the minority fraction with $\mu$SR~\cite{Majumder2018} does not appear to be consistent with the minority fraction observed in our measurements, as this line does not appear to broaden at low temperatures.

While Refs.~\cite{Shen2021,Veiga2019} attributed the suppression of magnetic susceptibility to dimerization (and/or tetramerization) of the material and presumably a spin-singlet ground state, our high-pressure relaxation rate measurements show that the system does not have a singlet ground state with gapped excitations or at least has a rather small value of gap $\Delta < 0.7$ m$e$V. The $1/T_1$ in fact follows a power-law temperature dependence $1/T_1~\propto T^\eta$, $\eta =1.7$ indicating the presence of gapless excitations. While it is not unheard of to discover such shallow power laws in ordered frustrated magnets~\cite{Yoshida2011,Quilliam2011Vesig}, when combined with a lack of magnetic order, it is tempting to attribute this behavior to the presence of a quantum spin liquid ground state with gapless spin excitations. It is also possible that the system is broken up into dimers and tetramers, but that the anisotropic interactions on those clusters lead to gapless excitations.

A number of quantum spin liquid models have been proposed to give rise to a Fermi surface of spinons~\cite{Motrunich2005,Lai2011,Barkeshli2013,He2018}. Given the $T$-linear Korringa relaxation in metals, previous works on spin liquid candidates have attributed a power-law temperature dependence with $\eta=1$~\cite{Dey2013} or slightly lower, for example $\eta = 0.8$~\cite{Gomilsek2017}, to a metal of spinon excitations. Meanwhile, magnons in conventionally ordered antiferromagnets ought to give rise to a much steeper power law, $T^3$ or higher. The intermediate temperature dependence ($\eta=1.7$) obtained here might be attributed to a spinon semi-metal with a nodal structure. For example, a $T^2$ ($\eta=2$) power-law relaxation in candidate quantum spin liquid materials 1T-TaS$_2$~\cite{Klanjsek2017} and EtMe$_3$Sb[Pd(dmit)$_2$]$_2$~\cite{Itou2010} were attributed to Fermionic spin excitations with a nodal gap. 

Indeed, theoretical work~\cite{OBrien2016} has shown that the ground state of the hyper-honeycomb Kitaev should consist of a nodal-line Majorana metal in zero field with Weyl nodes under applied field. It is not impossible that such a state occurs in the high-pressure phase of $\beta$-Li$_2$IrO$_3$. The two main problems with this picture are that, 1. it does not explain the suppression of magnetic susceptibility (shift) at quite high temperatures and 2. the real material is unlikely to exhibit a simple Kitaev model~\cite{Lee2015,Kim2015,Kim2016}. Nonetheless, the NMR results presented here indicate that $\beta$-Li$_2$IrO$_3$ in this intermediate pressure regime around 2 GPa, is not explained by a simple singlet state brought on by dimerization and/or tetramerization of the spin model and the novel power-law behavior of the relaxation rate calls for further investigation.

%
%
%
%
%

\section{Aknowledgements}
 
At the Université de Sherbrooke, this work was supported by the National Sciences and Engineering Research Council (NSERC), the Canada First Excellence Research Fund (CFREF) and the Fonds de Recherche du Québec : Nature et Technologies (FRQNT). We are grateful for technical support from S. Fortier and advice from A. Ataei and N. Doiron-Layraud. JAQ acknowledges financial support from the CNRS visiting scientist program and the Laboratoire d'excellence PALM at the U. Paris-Saclay,  as well as the SQM group for generously hosting his sabbatical year at the LPS. Work performed by JGA was supported by the U.S. Department of Energy, Office of Science, Basic Energy Sciences, Materials Sciences and Engineering Division under contract DE-AC02-05-CH11231 within the Quantum Materials program (KC2202).




\end{document}